\documentclass[epj]{webofc}
\usepackage[varg]{txfonts}   
\usepackage{tikz}
\wocname{EPJ Web of Conferences}
\woctitle{CONF12}
\definecolor{omavihrea}{rgb}{0.1 0.3 0.1}
\definecolor{omapun}{rgb}{0.5 0.0 0.15}
\definecolor{omasin}{rgb}{0.0 0.15 0.5}

\newcommand{\ptt}{p_{T}}
\newcommand{\ktt}{k_{T}}
\newcommand{\as}{\alpha_{\mathrm{s}}}
\newcommand{\ud}{\, \mathrm{d}}
\newcommand{\lqcd}{\Lambda_{\mathrm{QCD}}}
\newcommand{\qs}{Q_\mathrm{s}}
\newcommand{\xt}{{\mathbf{x}_T}}
\newcommand{\yt}{{\mathbf{y}_T}}
\newcommand{\ut}{{\mathbf{u}_T}}
\newcommand{\vt}{{\mathbf{v}_T}}
\newcommand{\rt}{{\mathbf{r}_T}}
\newcommand{\bt}{{\mathbf{b}_T}}

\newcommand{\ot}{{\mathbf{0}_T}}
\newcommand{\nabt}{\boldsymbol{\nabla}_T}
\newcommand{\nc}{{N_\mathrm{c}}}
\newcommand{\tr}{\, \mathrm{Tr} \, }
\newcommand{\half}{\frac{1}{2}}
\newcommand{\sigmadip}{{ \sigma_\textrm{dip} }}

\newcommand{\nr}[1]{(\ref{#1})}

\begin{document}
\selectlanguage{english}
\title{Initial conditions in AA and pA collisions}
%
%

\author{Tuomas Lappi\inst{1,2}\fnsep\thanks{\email{tuomas.v.v.lappi@jyu.fi}}
}

\institute{
Department of Physics, %
 P.O. Box 35, 40014 University of Jyv\"askyl\"a, Finland
\and
Helsinki Institute of Physics, P.O. Box 64, 00014 University of Helsinki,
Finland
}

\abstract{
A full understanding of the spacetime evolution of the QCD matter created in a heavy ion collision requires understanding the properties of the initial stages. In the weak coupling picture these are dominated by classical gluon fields, whose properties can also be studied via the scattering of dilute probes off a high energy hadron or nucleus. A particular challenge is understanding small systems, where LHC data is also showing signs of collective behavior. We discuss some recent results of on the initial matter production and thermalization in heavy ion collisions, in particular in the gluon saturation framework.
}
\maketitle

\section{Introduction}

Experimentally studying properties of the quark gluon plasma requires one to interpret measurements in relativistic heavy ion collisions. Relating these observables to the fundamental properties of QCD matter requires one to understand and quantitatively model the whole collision process, from the initial particle production up to the final freezeout. In this program, the initial stages are, by definition, least directly constrained by experiment. Their description must therefore rely most on QCD theory in the high energy limit; this will be the subject of this talk.

In the high collision energies reached in the LHC experiments one reaches the gluon saturation regime of QCD. In this regime the phase space density of small-$x$ gluons becomes nonperturbatively large even at semihard transverse momentum scales, i.e. scales where one can use weak coupling QCD methods. Here the CGC effective theory of QCD in the high energy limit allows one to organize the calculation in terms of a nonperturbatively large classical color field and quantum fluctuations around it~\cite{Lappi:2010ek,Gelis:2010nm}. In this talk we will first discuss how one can calculate properties of the initial state of the quark gluon plasma in this picture, using Classical Yang-Mills (CYM) simulations. We will then discuss some recent work on independently studying the structure of the classical field with dilute probes, where many of the needed cross sections are now being worked out to next-to-leading order accuracy in the QCD coupling. We finally discuss long range rapidity correlations in small (proton-proton or proton-nucleus) collision systems,  which could potentially provide a more direct experimental access to the  correlation structure of the color fields in the initial stage of the collision.

\section{From small-$x$ to the glasma initial state}

The bulk of the matter produced in the central rapidity region of a heavy ion collision originates from interactions of gluons that have only a small  fraction $x$ of the energy of the incoming nuclei. Gluons dominate over quarks because the strong logarithmic enhancement of the brehmsstrahlung spectrum: the probability to emit a gluon with $x\ll 1$ is proportional to $\as \ud x /x$. Thus to leading order in the weak coupling the distribution of initial matter production is independent of rapidity $y = \ln 1/x$. The experimentally observed approximate boost-invariance of the particle multiplicity is therefore natural in a weak coupling picture. 

When the total collision energy is very large, there is enough phase space for the emitted gluons to themselves act as sources for a further emission of even smaller-$x$ gluons. This leads to an exponentially developing cascade of gluon emissions, where every subsequent step in the emission process is proportional to an additional power of $\as \ln s$. At high enough energy $s$ this cascade of gluon emissions needs to be resummed, because the large energy logarithm compesates for the smallness of the coupling. The result of this resummation is that  the produced gluon multiplicity is proportional to a small power $\sim \as$ of the total center of mass energy: again a trend in agreement with the experimental observations. Making these statements more quantitative requires an actual procedure for performing this resummation of the large energy logarithms: this is performed with high energy renormalization group equations known by the acronyms BFKL, BK or JIMWLK.

The BFKL equation resums the gluon cascade as a perturbative process, including the emissions of soft gluons to all orders in $\as$ in the leading logarithmic limit. It is a linear equation that indeed leads to a solution that grows exponentially with rapidity. However, eventually this approach will lead to nonperturbatively large occupation numbers for gluon states in the wavefunction. When this happens, the perturbative approximation  ceases to be valid. Gluon mergings become parametrically equally important as the splittings, and the linear BFKL high energy renormalization group equation must be replaced with a nonlinear BK or JIMWLK one. This phenomenon is known as gluon saturation.

The occupation numbers of gluons in the high energy nuclear wavefunction always decrease as a function of the transverse momentum $\ktt$, reverting to the linear regime at large $\ktt$. The intrinsic transverse momentum scale at which the nonlinearities become important is known as the saturation scale $\qs$. When the energy is high enough, the QCD coupling at the scale $\qs$ is weak. At the saturation scale the two terms of the Yang-Mills covariant derivative $\partial_\mu + igA_\mu$ are of the same order because $A_\mu \sim 1/g$, signaling a breakdown of the perturbative expansion. This leads to the typical gluon saturation power counting, where occupation numbers are nonperturbatively large $f(k) \sim A_\mu A_\mu \sim 1/\as$ at the scale $k\sim \qs \gg \lqcd$ so that $\as \ll 1$. A gluonic system satisfying these conditions is, to leading order in $\as$, a \emph{classical color field}. The question of whether this weak coupling, but nonperturbative, saturation picture is relevant for a particular kinematical regime can always be debated; this cannot be resolved within the (self consistent) classical approximation alone. It is however clear that the picture becomes better and better justified when the collision energy $\sqrt{s}$ increases.

In the saturation regime the number of gluons in the nucleus, i.e. the gluon distribution, is in fact not the most convenient degree of freedom to describe the target. In stead, in practical calculations are performed in terms of the \emph{Wilson line}, which for a left-moving nucleus is written as
\begin{equation}\label{eq:wline}
V(\xt) = P \exp \left\{i g \int  \ud x^+ 
A^-(\xt,x^-)\right\} \quad  \in \quad \textnormal{SU(3),}
\end{equation}
a path-ordered exponential in the target color field along the eikonal trajectory of a high energy probe advancing along the positive $z$ axis.  The Wilson line depends on the transverse coordinate. The light cone time coordinate $x^+$ is integrated over following the trajectory of a probe particle. The dependence on the light cone longitudinal coordinate $x^-$ is insignificant in the high energy limit, since a probe with a large $p^+$ momentum interacts with the target instantaneously in $x^-$.

\begin{figure}
\sidecaption
\centering 
\resizebox{0.5\textwidth}{!}{
\includegraphics[width=4.1cm]{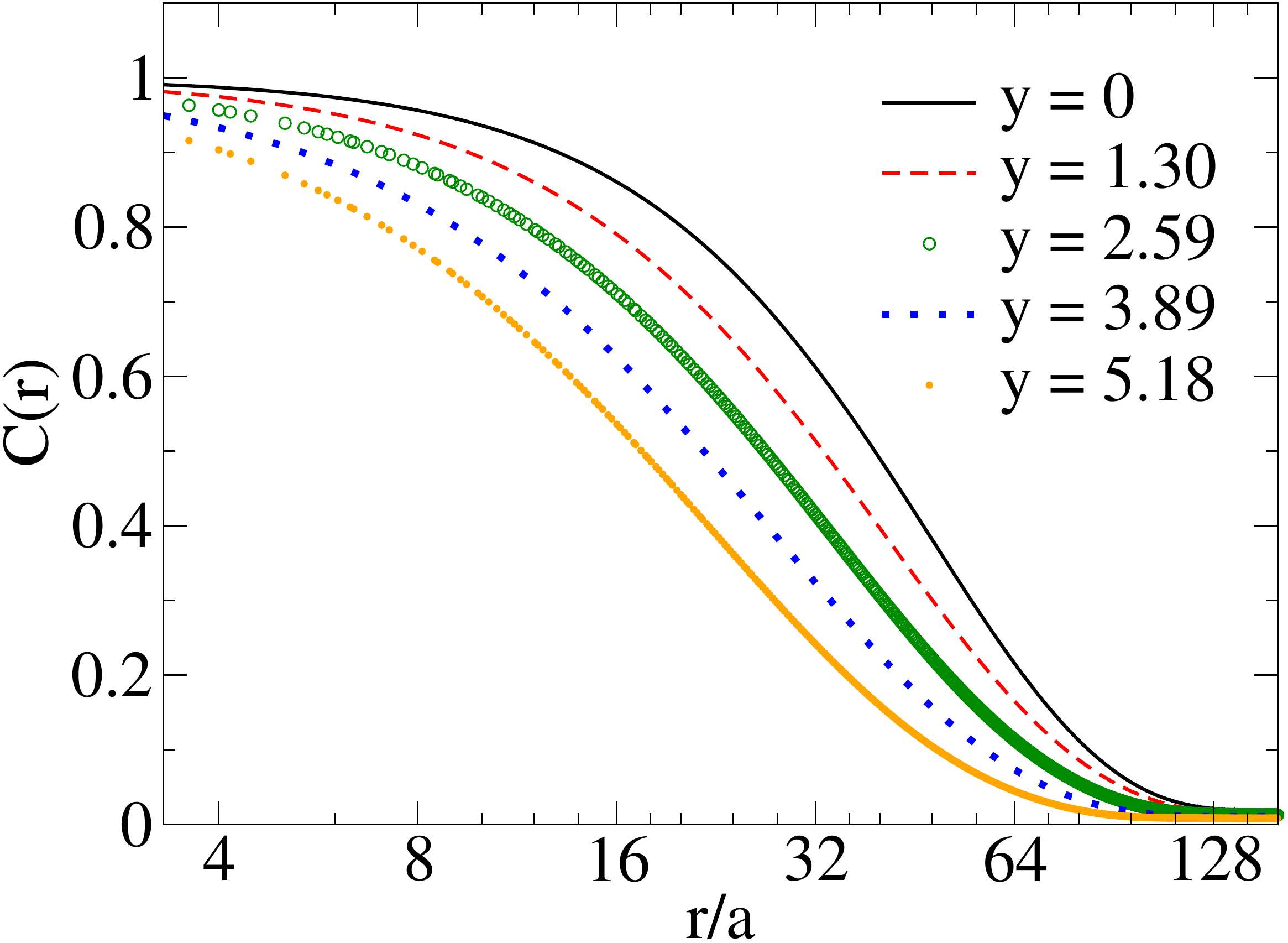}
\begin{tikzpicture}[overlay]
\draw[omapun,line width=1pt,dashed] (-3.7,1.81) -- (-1.5,1.81);
\draw[omasin,line width=1pt,->,dashed] (-1.7,1.8) -- (-1.7,0.35);
\draw[omasin,line width=1pt,->,dashed] (-2.5,1.8) -- (-2.5,0.35);
\end{tikzpicture}
}
\caption{Wilson line correlator \nr{eq:defqs} as a function of transverse coordinate, from the JIMWLK simulations~\cite{Lappi:2011ju}. The dashed lines indicate the saturation scale as defined by the condition \nr{eq:defqs}.}
\label{fig:wlinecorr}
\end{figure}
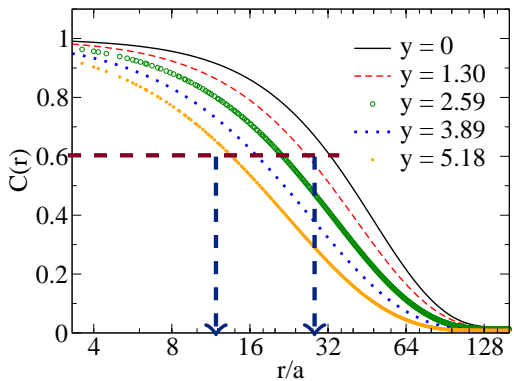

Expressed in terms of the Wilson lines, the saturation scale  $\qs$ is the inverse of their correlation length in the transverse plane. A gauge invariant definition can be obtained using the expectation  value of the dipole operator $\tr V^\dag(\xt) V(\yt)$, which interpolates between $\nc$ at $\xt=\yt$  and zero at large distances, when the Wilson lines are completely uncorrelated. For a precise definition, one must choose some limiting value between $\nc$ and 0; for concreteness one often takes
\begin{equation}\label{eq:defqs}
 \frac{1}{\nc} \left< \tr V^\dag(\ot)V(\xt) \right> 
= e^{-\half}  
\Longleftrightarrow \xt^2 = \frac{2}{\qs^2}.
\end{equation}
Figure \ref{fig:wlinecorr} demonstrates the extraction of this characteristic length from a numerical  solution~\cite{Lappi:2011ju} of the JIMWLK renormalization group equation.

The Wilson line is formed from the covariant gauge color field, whose largest component for a color charge with a large momentum $p^\pm$ is $A^\pm$. In order to have an interpretation of this color field in terms of partons one must gauge transform this field into the light cone gauge. The result is a transverse pure gauge field
\begin{equation}\label{eq:trpureg}
 A^i = \frac{i}{g} V(\xt) \partial_i V^\dag(\xt).
\end{equation}
While the covariant gauge field lives on the light cone of the corresponding source, $\sim \delta(x^\mp)$ for a color current in the $\pm$-direction, the transverse pure gauge field is delocalized in the longitudinal coordinate, $\sim \theta(x^\mp)$. Physically this is the spacetime picture~\cite{Lappi:2006fp} that can be interpreted in terms of the Heisenberg uncertainty principle: the classical field corresponds to small-$x$ degrees of freedom, i.e. gluons with a \emph{small} $p^\pm$, it is therefore delocalized in $x^\mp$. 

\begin{figure}
\sidecaption
\centering
\includegraphics[width=0.5\textwidth]{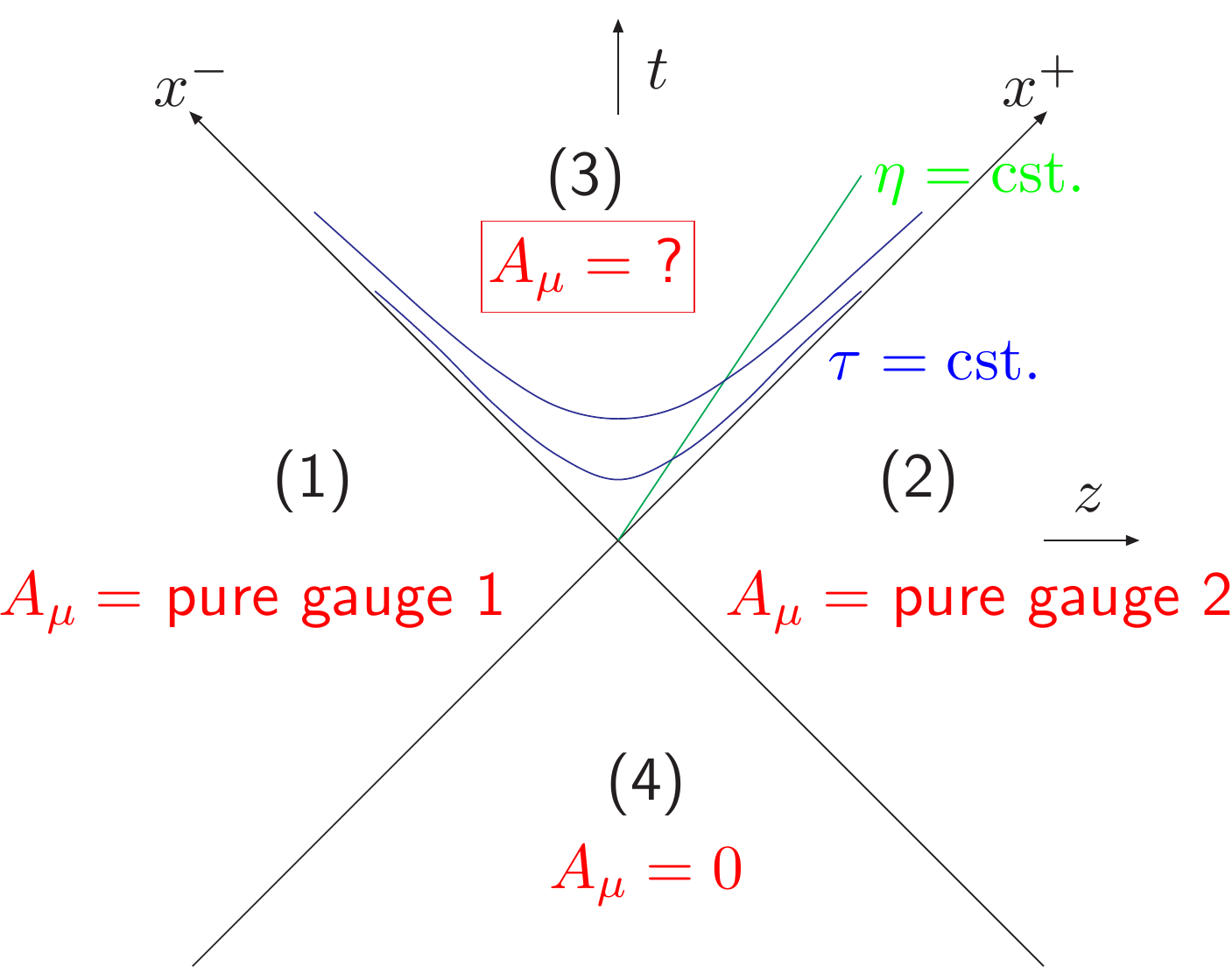} 
\caption{The classical gauge field in the collision of color charged systems. The fields in the regions (1) and (2) are transverse pure gauges, Eq.~\nr{eq:trpureg}. The initial condition for the nontrivial field inside the future light cone, $x^+>0,x^->0$ is given by Eq.~\nr{eq:ic}.
}
\label{fig:spacet}
\end{figure}

The light cone gauge field is the basis of using the classical field picture to calculate the color field in a heavy ion collision. One starts from separate independent pure gauge fields, Eq.~\nr{eq:trpureg}, for the two colliding nuclei (1) and (2). The transverse pure gauge field is gauge equivalent to the vacuum outside of the light cone, i.e. it carries no energy density. However, the superposition of two such fields from independent sources is not any more a pure gauge. A straightforward calculation~\cite{Kovner:1995ja,Kovner:1995ts} in fact allows one to calculate the field at $\tau=\sqrt{2x^+x^-}=0^+$ after a collision of two sheets of color glass as
\begin{equation}
\label{eq:ic}
\left. A^i\right|_{\tau=0} = A^i_\textnormal{(1)} + A^i_\textnormal{(2)} 
\quad 
\textnormal{and}
\quad 
\left. A^\eta\right|_{\tau=0} = \frac{ig}{2} [A^i_\textnormal{(1)} , A^i_\textnormal{(2)} ],
\end{equation}
where we are working in the gauge $A_\tau=0$ with the spacetime rapidity $\eta = \half \ln x^+/x^-$ as the longitudinal coordinate. This spacetime structure is illustrated in Fig.~\ref{fig:spacet}. The field inside the future light cone is referred to as the \emph{glasma} field~\cite{Lappi:2006fp}.

No analytical solution for the Yang-Mills equation of motion for the glasma field is known, but based on the initial condition one can easily deduce some of its most important features (see also ~\cite{Lappi:2006hq,Chen:2013ksa,Chen:2015wia}. Precisely at $\tau=0$ the field consists of purely  longitudinal chromoelectric and -magnetic fields. They vary in the transverse direction at the typical length scale $1/\qs$, and could thus be thought of equivalently as gluons of momentum $\sim \qs$ or color strings of size $1/\qs$. The interpretation of $1/\qs$ as a \emph{correlation length} is crucial for understanding the effect of the glasma fields on multiparticle correlations. After a time $\sim 1/\qs$ the field modes decohere, and the energy density consists in equal amounts of longitudinal  and transverse components $E_z^2 \sim B_z^2 \sim B_x^2+B_y^2 \sim E_x^2+E_y^2$. Translating these into diagonal components of the energy momentum tensor one finds that precisely at $\tau=0$ the system has a very strong negative vacuum-energy-like longitudinal pressure $P_L \sim -\varepsilon \sim -P_\perp$. After the decoherence time $\tau\sim 1/\qs$ it becomes very strongly anisotropic, corresponding to gluons with momenta only in the transverse directions, $P_\perp \sim 1/\tau \gg P_L$. This is the starting point of the initial state isotropization problem in heavy ion collisions, great progress on which has recently been made in a kinetic theory framework going beyon the classical field limit~\cite{Kurkela:2015qoa}.

\begin{figure}
 \includegraphics[width=0.49\textwidth,clip=true]{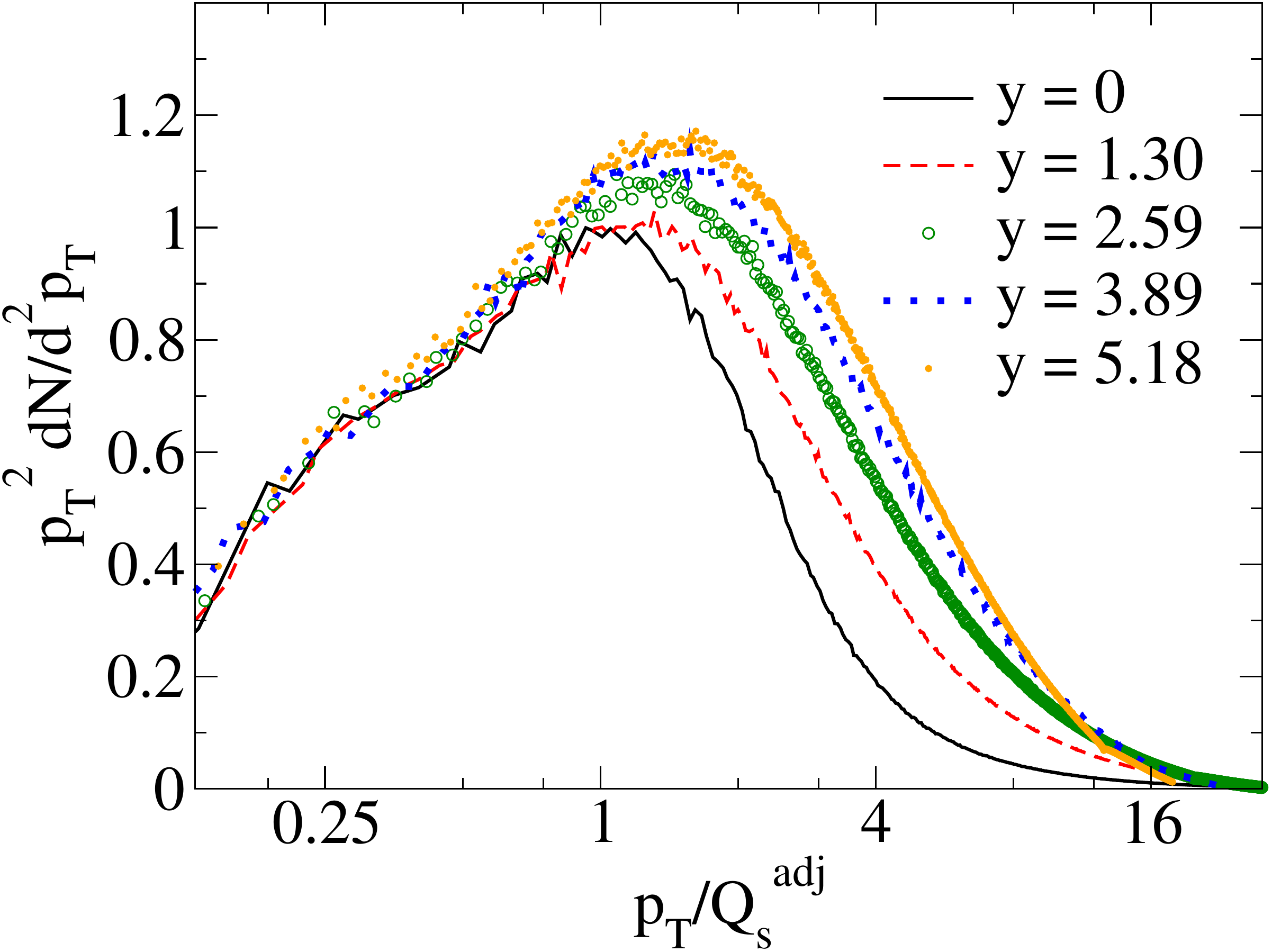}
\includegraphics[width=0.49\textwidth] {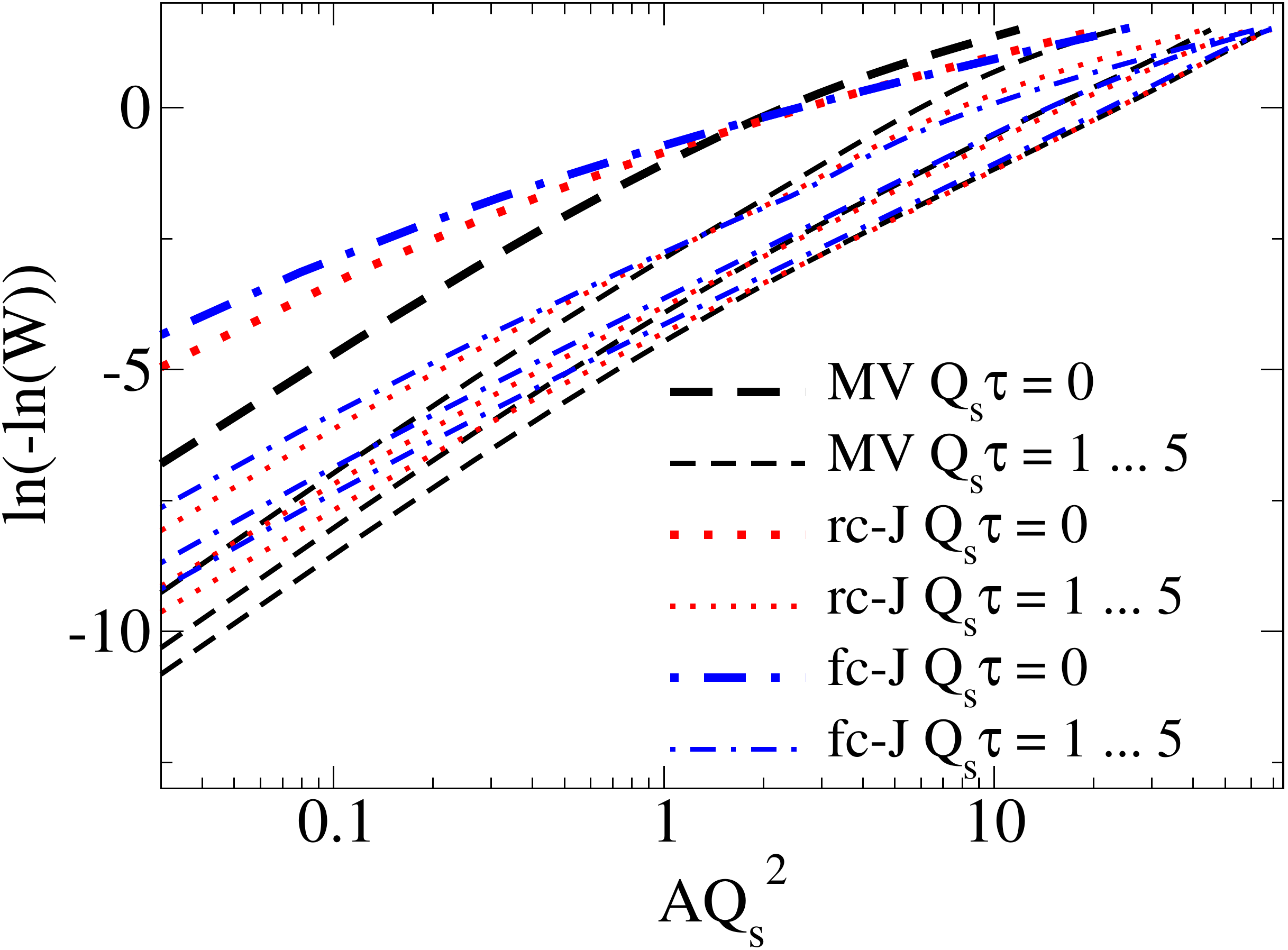}
\caption{Left: gluon spectrum in the glasma expressed in terms of the scaling variable $\ptt/\qs$~\cite{Lappi:2011ju}. The small momentum part of the spectrum collapses to an universal curve. Right: area-dependence of the spatial Wilson loop $W(A)$~\cite{Dumitru:2014nka} shows the same collapse to a ($\tau$-dependent) universal value for large areas $A$.}
\label{fig:universality}
\end{figure}

The nature of the saturation scale as the dominant characteristic scale in the problem carries over from the wavefunction of one nucleus to the matter produced in the collision of two. By Fourier-decomposing the classical field into momentum modes, one can calculate the corresponding gluon spectrum, see Fig.~\ref{fig:universality}. For large momenta $\ptt \gg \qs$ it reduces to a perturbatively calculable spectrum whose form depends on the initial condition, i.e. on the distribution of Wilson lines. For small momenta $\ptt \ll \qs$ the gluon spectrum collapses to a universal form close to a classical thermal distribution $\sim 1/\ptt$. An alternative manifestly gauge invariant way of studing the same universality is to look at the area-dependence of a spatial Wilson loop in the transverse plane~\cite{Dumitru:2014nka}:
\begin{equation}
 W(A) = \tr \mathop{P} \exp\left\{i g \oint_{\partial A} \ud x^\mu A_\mu\right\}.
\end{equation}
Here a very similar phenomenon is observed: for small loops $A\ll 1/\qs^2$ the area-dependence depends on the details of the distribution of Wilson lines, but for large loops both the magnitude and the area dependence are universal (see Fig.~\ref{fig:universality}). This universality due to nonlinear gluon interactions is a manifestation of gluon saturation in the final state.

\section{Control measurements}

As we have discussed, the properties of the glasma fields can be calculated once one knows the Wilson lines \nr{eq:wline} describing the color fields of the colliding nuclei. To make the glasma picture of the initial stage consistent, controlled and quantitative, one must be able to probe these same Wilson lines separately in control experiments. This means collisions of small, dilute, well-controlled probes off a large nucleus.

\begin{figure}
\sidecaption
\centering
\includegraphics[width=0.5\textwidth]{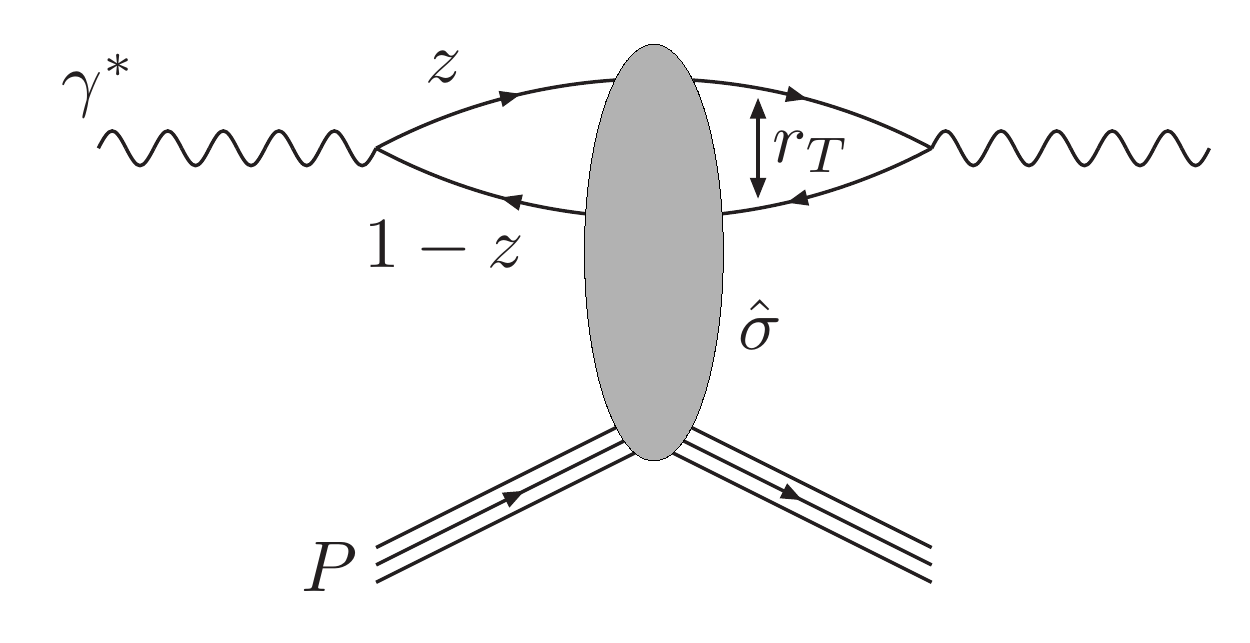}
\caption{ Deep inelastic scattering, i.e. $\gamma^*$-target collision, in the dipole picture. The virtual photon splits into a quark-antiquark pair, which interacts eikonally with the target color field. Here one is interested in the total cross section which, via the optical theorem, is related to the imaginary part of the forward elastic scattering amplitude. Thus in the final state the dipole must recombine to give the same virtual photon state.}
\label{fig:dipole} 
\end{figure}

The simplest such process is deep inelastic scattering (DIS). To connect it to the CGC framework it is convenient to think of the scattering process in the ``dipole picture.'' Here a  virtual photon $\gamma^*$ fluctuates into a quark-antiquark pair, which then eikonally interacts with the target color field, see Fig.~\ref{fig:dipole}. The cross section can correspondingly be factorized as
 \begin{equation}
\sigma^{\gamma^*H}_{\textnormal{tot}} = 
\int \ud^2\rt \ud z
\left| \Psi_{\gamma^* \to q\bar{q}} (\rt,z)\right|^2 
 \sigmadip  (\rt,z),
 \end{equation}
where the QED light cone wave function $\Psi_{\gamma^* \to q\bar{q}}$ quantifies the probability for the virtual photon to split into a dipole of size $r$, enforcing $r\sim 1/Q$. The QCD dynamics is parametrized by the dipole cross section $ \sigmadip$, which is a nothing but our previous Wilson line correlation function
\begin{equation}
\sigmadip(\rt) = \int \ud^2 \bt 
\frac{1}{\nc}
\mathrm{Tr}
\left\langle 1 -
V^\dag\left(\bt + \frac{\rt}{2}\right)V\left(\bt - \frac{\rt}{2}\right) \right\rangle.
\end{equation}

The high energy renormalization group equations (BK or JIMWLK) predict the dependence of the dipole cross section on energy (or equivalently Bjorken $x$ in the case of DIS). Although the evolution equations are derived in weak coupling QCD, they still need a nonperturbative initial condition that needs to be fit to experimental data. Thus a good CGC description of the initial stages of a heavy ion collision should be consistent with HERA measurements, and give testable predictions for future nuclear DIS experiments, e.g. at the EIC~\cite{Accardi:2012qut}. The agreement with HERA data has indeed been demonstrated in several recent calculations, see e.g. \cite{Rezaeian:2012ji,Lappi:2013zma}.

Another important control measurement is to use another dilute, relatively well understood, system of quarks and gluons as a probe, namely a proton at relatively large $x$. This is done by looking at single particle production and multiparticle correlations at forward rapidity in proton-nucleus collisions. Here the particle production cross section is proportional to the nuclear \emph{unintegrated} gluon distribution, which is given simply by a Fourier-transform of the dipole cross section. Leading order calculations with up-to-date dipole cross sections now give a relatively good description of the spectra, albeit with the normalization corrected by a ``$K$-factor''~\cite{Albacete:2010bs,Albacete:2012xq,Rezaeian:2012ye,Lappi:2013zma}. This normalization uncertainty cancels out in $R_{pA}$, the ratio of proton-nucleus and proton-proton cross sections, corrected by the nuclear geometry. Since other uncertainties in $R_{pA}$ are rather small, it is important to treat the nuclear geometry correctly also in the theory calculations, indeed the largest difference between earlier~\cite{Albacete:2010bs} and more recent~\cite{Lappi:2013zma} calculations is precisely here. In a similar way the importance of treating carefully the nuclear geometry has been seen in calculations of $J/\Psi$ production at forward rapidity~\cite{Fujii:2013gxa,Ducloue:2015gfa}.

More recently attention on the theory side has turned to carrying out the program of NLO calculations of dilute-dense collision systems to next-to-leading order (NLO) accuracy.  The leading order Balitsky-Kovchegov (BK)~\cite{Balitsky:1995ub,Kovchegov:1999yj,Kovchegov:1999ua} equation is now routinely used in phenomenology. Also the NLO version of the equation was derived some time ago~\cite{Balitsky:2008zza}, but no serious attempts to solve it were undertaken for a long time. The equation as originally derived is unstable~\cite{Lappi:2015fma} due to large transverse momentum logarithms. Recently a way to resum these logarithms was developed~\cite{Iancu:2015vea,Iancu:2015joa}, leading to a stable solution of the whole equation at NLO accuracy. It turns out~\cite{Lappi:2016fmu} that with a suitable choice of a constant under the log, one can arrange so that most of the NLO corrections are included in the resummation and the remaining finite NLO terms are small. A first fit of HERA DIS data in this framework has been performed~\cite{Iancu:2015joa}, and the NLO equation could be expected to become a frequently used tool in high energy QCD.

Forward single inclusive particle production in proton-nucleus collisions has similarly been the focus of much attention recently. The cross section was calculated to NLO order in an important calculation a few years ago~\cite{Chirilli:2011km,Chirilli:2012jd}. First numerical evaluations of the cross section~\cite{Stasto:2013cha} led to the result that the NLO corrections cause the cross section to turn negative at large transverse momenta. Several interpretations of this ``negativity problem'' have been presented in the subsequent literature~\cite{Kang:2014lha,Altinoluk:2014eka,Watanabe:2015tja}. The negativity problem can be traced back to the way the small-$x$ divergence is subtracted from the cross section in order to be absorbed into the BK evolution of the target~\cite{Ducloue:2016shw}. A physical feature in these discussions is the need to impose ordering in the longitudinal momentum in the target ($k^-$, if the probe is moving in the positive $z$ direction) in stead of the probe $k^+$, which would be the more straightforward variable in the cross section calculation. Imposing this ordering in target longitudinal momentum seems to significantly alleviate the negativity problem. More recently a reformulation of the cross section that evokes an exact  integral representation of the BK equation and makes the cross section explicitly positive was presented~\cite{Iancu:2016vyg}. This proposal has  yet to be implemented in a quantitative numerical  calculation. 

\section{Correlations}

Let us now move back to denser collision systems for a short discussion of long range rapidity correlations. 
A very generic causality argument~\cite{Dumitru:2008wn} shows that correlations between very far away rapidities can only originate at a very early stage of the collision. This is completely analogous to the way large scale structrures in the cosmological microwave background reveal information about the earliest stages of the universe, because the correlations can only originate at a time when the regions were in causal contact.  Thus long range rapidity correlations have the potential to be directly sensitive to the initial stage of the collision.

The most analyzed multiparticle correlations, usually one that is long range in rapidity, in a heavy ion collision are $v_n$'s, i.e. the Fourier series coefficients of the azimuthal anisotropy in the produced particle spectra. These ``flow coefficients'' are traditionally presented as the Fourier-coefficients of the single particle distribution with respect to a reaction plane determined by the impact parameter direction between the colliding nuclei. The impact parameter is, however, not actually measured experimentally. In stead, the impact parameter direction is determined from the other particles produced in the event, often with a rapidity separation to the ``single'' particle. This makes a $v_n$ measurement always a multiparticle correlation one, which is even more explicit in more recent (cumulant) analysis methods (see e.g. Ref.~\cite{Chatrchyan:2013nka}).

The geometry of the collision system is the ultimate infinite-range rapidity correlation: particle production at all rapidities is sensitive  to the same positions of the incoming nucleons in the nucleus. Geometry  gives rise to a coordinate space azimuthal asymmetry, but in most particle production scenarios the momentum distribution is initially azimuthally symmetric. 
 If interacting matter is produced in the collision and lives for a long enough time, collective interactions in the form of work done by pressure gradients can transform the initial coordinate space asymmetry into an asymmetry in the local fluid velocity, i.e. the momentum distribution of the particles. This is the conventional hydrodynamical explanation of anisotropic azimuthal flow, and we have every reason to believe that it is the correct interpretation in the case of large collision systems, i.e. nucleus-nucleus collisions. 

For parametrically smaller systems, where the size in the transverse plane is not much larger than the correlation area  of the color fields $1/\qs^2$, there is an alternative mechanism that can generate correlations between the produced particles directly in momentum space, unrelated to the overall coordinate space geometry of the collision system.
Although the mechanism is present also in a ``dense-dense'' collision system, it is more straightforward to explain the effect in the case of a dilute probe of the target color field. Here the physical picture of particle production is the following~\cite{Kovner:2010xk,Lappi:2015vta}. One starts from a collinear high-$x$ quark or gluon from the probe hadron. To produce hadrons it will scatter from the target color field, getting a transverse momentum kick $\sim \qs$ from the transverse chromoelectric  field of the target. Now the target consists of domains of color fields of typical size $\sim 1/\qs^2$. Thus two produced particles will be correlated if they hit the same domain. This is very likely if the overlap area $S_\perp$ of the probe and target is not much larger than $1/\qs^2$, otherwise the correlation is washed out by the increasing likelihood that the particles are produced from independent domains. In order to get a momentum kick in the same direction from the target color field domain in a particular color state, the two incoming particles have to have the same color. Combining these two effects we see that the correlation is suppressed by a factor $1/(S_\perp \qs^2 \nc^2)$ compared to uncorrelated particle production. Thus this is an effect that is parametrically larger for smaller collision systems, contrary to the usual final state collective flow-induced correlation, which requires a long lifetime and thus a large volume.

The same physical mechanism is behind several recent calculations of azimuthal anisotropies in the CGC framework. Different approximation strategies are most straightforwardly discussed in terms of the Wilson line
$V(\xt)$, which can be parametrized in terms of a color charge density $\rho$ as:
\begin{equation}
 V(\xt) 
= P \exp\left\{i g \int \ud x^- \frac{\rho(\xt,x^-)}{\nabt^2} \right\}, 
\end{equation}
For a calculation of two-particle correlations one needs at some level to calculate correlators of four Wilson lines such as
\begin{equation}\label{eq:dipsqr}
 \left<\tr V^\dag(\xt) V(\yt) \tr V^\dag(\ut) V(\vt) \right>,
\end{equation}
which brings a sensitivity to the statistical properties  of  the Wilson lines beyond the two-point function used for single inclusive particle production.

The ``ridge'' correlation was calculated in Ref.~\cite{Dusling:2013oia} in terms of the ``glasma graph'' approximation. This is a $\ktt$-factorized approximation  which can be derived by linearizing the Wilson lines in terms of the color charge $\rho$, and calculating higher point functions such as \nr{eq:dipsqr} by assuming that the correlators of $\rho$ are Gaussian. The property of Gaussianity allows one to express all higher point operators in terms of the two point function of $\rho$'s: in the usual $\ktt$-factorized approximation this is then again replaced by the full nonlinear two-point function of Wilson lines that satisfies the BK equation.
The ``electric field domain model'' of \cite{Dumitru:2014dra} investigated the effect of an additional intrinsically non-Gaussian correlation between the color charges $\rho$ that has not been seen in the usual JIMWLK/BK setup. The dilute-dense case with a full JIMWLK evolution was studied in Ref.~\cite{Lappi:2015vha}, treating the probe in the dilute collinear approximation, but without linearizing in the color charge $\rho$. The CYM calculation of~\cite{Schenke:2015aqa} used the MV model, but includes fully the nonlinear color field, supplemented with a CYM evolution that introduced also non-equilibrium final state correlations.
We stress that the main physics idea in these calculations is the same, only approximations in treating the nonlinearities differ. In a  later study~\cite{Lappi:2015vta} we compared the differences of these approximations in an apples way in the dilute-dense limit. The ``nonlinear Gaussian'' approximation where the $\rho$'s are assumed to have Gaussian correlators, but the nonlinear relation between them and the Wilson line $V(\xt)$ is treated to all orders, was found to be accurate at least within 10\%. The difference between the full JIMWLK result and the ``Glasma graph'' calculation that is linearized in $\rho$ at intermediate stages differs from the full result by at most a factor of 2 (for the coefficient $v_2$) and in most cases less.

\section{Conclusions}

In conclusion, we have here given a brief overview of the connection between the CGC picture of the small-$x$ degrees of freedom in a high energy hadron or nucleus, and the initial stage of a relativistic heavy ion collision. We have seen that this is characterized by the concept of a nonperturbatively strong classical gauge field, which leads to an anisotropic system of gluons in the initial stage of quark gluon plasma formation. We have then discussed recent advances in calculating at NLO accuracy cross sections for processes where this strong color field is probed by a dilute probe. Following through the full calculation of the glasma field to NLO accuracy is yet a more challenging undertaking for the future~\cite{Gelis:2008rw}. Finally we discussed azimuthal multiparticle correlations, usually parametrized in terms of the flow coefficients $v_n$ in small collision systems. Here there is likely an intreresting interplay between initial state color field and final state hydrodynamical correlations, which still remains to be fully understood quantitatively.

 This work has been  supported by the Academy of Finland, projects 267321,  273464 and  303756 and  by the European Research Council, grant ERC-2015-CoG-681707.

%
%

%
%
%

\bibliography{spires}

\end{document}